\documentclass[12pt,a4paper,eqsecnum,showkeys,showpacs,aps]{revtex4}
\usepackage{graphicx,amsfonts,amsmath,amssymb}
\providecommand{\xii}{\xi\hspace{-1.5mm}
\raisebox{-1.5ex}{\textit{\scriptsize i}}\hspace{0.5mm}{}}
\providecommand{\xione}{\xi\hspace{-1.5mm}
\raisebox{-1.5ex}{{\scriptsize 1}}\hspace{0.2mm}{}}
\providecommand{\xitwo}{\xi\hspace{-1.5mm}
\raisebox{-1.5ex}{{\scriptsize 2}}\hspace{0.2mm}{}}
\providecommand{\xibartwo}{\bar{\xi}{}\hspace{-1.5mm}
\raisebox{-1.5ex}{{\scriptsize 2}}\hspace{0.2mm}{}}
\providecommand{\xibari}{\bar{\xi}{}\hspace{-1.5mm}
\raisebox{-1.5ex}{\textit{\scriptsize i}}\hspace{0.2mm}{}}
\providecommand{\thetaj}{\theta\hspace{-2.0mm}
\raisebox{-1.5ex}{\textit{\scriptsize j}}\hspace{1.0mm}{}}
\providecommand{\thetaone}{\theta\hspace{-1.5mm}
\raisebox{-1.5ex}{{\scriptsize 1}}\hspace{0.2mm}{}}
\providecommand{\thetatwo}{\theta\hspace{-1.5mm}
\raisebox{-1.5ex}{{\scriptsize 2}}\hspace{0.2mm}{}}
\providecommand{\thetabarone}{\bar{\theta}{}\hspace{-1.5mm}
\raisebox{-1.5ex}{{\scriptsize 1}}\hspace{0.2mm}{}}
\providecommand{\thetabartwo}{\bar{\theta}{}\hspace{-1.5mm}
\raisebox{-1.5ex}{{\scriptsize 2}}\hspace{0.2mm}{}}
\begin{document}
\title{On group properties and reality conditions \\%
       of UOSp(1$|$2) gauge transformations}%
\author{Kostyantyn Ilyenko}%
\email{kost@ire.kharkov.ua}%
\affiliation{Institute for Radiophysics and Electronics of NAS of
             Ukraine, vul.~Akad.~Proskury, 12, Kharkiv, 61085, Ukraine}%
\date{November 10, 2008}%
%
\begin{abstract}
For \textit{osp}(1$|$2;~$\mathbb{C}$) graded Lie algebra, which
proper Lie subalgebra is \textit{su}(2), we consider the
Baker-Campbell-Hausdorff formula and formulate a reality condition
for the Grassmann-odd transformation parameters that multiply the
pair of odd generators of the graded Lie algebra. Utilization of
\textit{su}(2)-spinors clarifies the nature of Grassmann-odd
transformation parameters and allow us an investigation of the
corresponding infinitesimal gauge transformations. We also explore
action of the corresponding group element of UOSp(1$|$2) on an
appropriately graded representation space and find that the graded
generalization of hermitian conjugation is compatible with the
Dirac adjoint. Consistency of generalized (graded) unitary
condition with the proposed reality condition is shown.
\end{abstract}
\maketitle

\newpage
\section{Introduction}

A natural extension of the Lie algebras, which underlie the modern
gauge theory, are graded Lie algebras introduced and studied to
some extent, for example, in the
articles~\cite{Berezin1970,CNS1975,Kac1977}. In this paper we
study those properties of the graded extension,
\textit{osp}(1$|$2; $\mathbb{C}$), of \textit{su}(2) Lie algebra,
which are pivotal for the purposes of constructing a meaningful
gauge theory of the Yang-Mills type (see, e.g.,
\cite{Neeman1979,BL1996,Ilyenko2001}). The explicit form of
\textit{osp}(1$|$2;~$\mathbb{C}$) defining relations (written as
in the articles~\cite{BL1996,Ilyenko2001,SW2005}) utilizes the
Pauli matrices and strongly suggests a relation to spinors.
Exponentiating the algebra to obtain the graded Lie group
UOSp(1$|$2), we observe the necessity of introduction of
anticommuting (Grassmann-odd) spinors, which multiply the odd
generators of the graded Lie algebra. We study some of the
infinitesimal properties of composition law of the group
transformations and show how to formulate a reality condition for
the Grassmann-odd spinors. Action of the corresponding group
element of UOSp(1$|$2) on an appropriately graded representation
space is explored and consistency of the graded generalization of
hermitian conjugation with the Dirac adjoint is demonstrated.
Finally, for the example of Grassmann algebra on two generators
(generalization to the case of even or infinite number of
generators of Grassmann algebra is straightforward), we show that
the reality condition is compatible with the proper generalization
of unitary condition~\cite{SW2005}, thus, making necessary
preparations for an investigation of the gauge invariance of the
proposed field strength~\cite{BL1996,Ilyenko2001} for such a
graded Yang-Mills theory.

\section{Graded Lie algebra \textit{osp}(1$|$2;~$\mathbb{C}$)}
\label{AlgGr}%

The algebra \textit{osp}(1$|$2;~$\mathbb{C}$) is a graded
extension of \textit{su}(2) by a pair of odd generators,
$\tau_{A}$, which anticommute with one another and commute with
the three even generators, $T_{a}$, of \textit{su}(2). It is
customary to assign a degree, deg\,$T_\alpha$, to the even
(deg\,$T_a$ = 0) and odd (deg\,$\tau_A$ = 1) generators. We use
the square brackets to denote the commutator and the curly ones to
denote the anticommutator. The defining relations have the
form~\cite{BL1996,Ilyenko2001,SW2005} (also
cf.~\cite{Kac1977,SNR19772,Hughes1981}):
\begin{eqnarray}
&[T_{a}, T_{b}] = i\varepsilon_{abc}T^{c},\;\;\; [T_{a}, \tau_{A}]
= \dfrac{1}{2}\,(\sigma_a)_{A}^{\hphantom{A}B}\tau_{B},& \nonumber \\
&\{\tau_{A}, \tau_{B}\} = \dfrac{i}{2}\,(\sigma^a)_{AB}T_{a}.&
\label{DefRel}
\end{eqnarray}
Summation is assumed over all repeated indices. Lowercase Roman
indices from the beginning of the alphabet run from 1 to 3;
uppercase Roman indices run over 1 and 2; $\delta_{ab}$ =
$\delta^{ab}$ ($\delta_{ab}$ = $\delta_{ba}$), $\varepsilon_{abc}$
($\varepsilon_{123}$ = $\varepsilon^{123}$ = 1) and
$\epsilon_{AB}$ ($\epsilon_{12}$ = $\epsilon^{12}$ = 1) are the
three dimensional identity matrix and the Levi-Civita totally
antisymmetric symbols in three and two dimensions, respectively;
the matrices $(\sigma_a)_{A}^{\hphantom{A}B}$ [$(\sigma^a)_{BA}$ =
$(\sigma^a)_{AB}$ = $\delta^{ab}(\sigma_b)_{AB}$ =
$\delta^{ab}(\sigma_b)_{A}^{\hphantom{A}C}\epsilon_{CB}$] are just
the usual Pauli matrices:%
\begin{eqnarray}
&\!\!\!(\sigma^a)_{A}^{\hphantom{A}B} =
(\sigma_a)_{A}^{\hphantom{A}B} = \left[ \left(
\begin{array}{cc}
0 & 1 \\ 1 & 0
\end{array}
\right), \left(
\begin{array}{rl}
 0 & i \\
-i & 0
\end{array} \right),%
\left( \begin{array}{cc} 1 &\! 0 \\ 0 &\! -1
\end{array}
\right) \right], & \nonumber \\%
&\!\!\!(\sigma^a)_{AB} = (\sigma_a)_{AB} = \left[ \left(
\begin{array}{cc}
\!\! -1 & 0 \\
      0 & 1
\end{array}
\right), \left(
\begin{array}{rl}
-i &       0 \\
 0 & \!\! -i
\end{array} \right),%
\left( \begin{array}{cc}%
 0 &\! 1 \\%
 1 &\! 0
\end{array}
\right) \right]. & \nonumber 
\label{PauliM}
\end{eqnarray}
We use the Levi-Civita symbols in two dimensions to raise and
lower uppercase Roman indices paying attention to their
antisymmetric properties:
\begin{displaymath}
\Sigma = \|\epsilon^{AB}\| = \left(
\begin{array}{rr}
0 &  1 \\ -1 & 0
\end{array}
\right) = \|\epsilon_{AB}\| = -\Sigma^{-1}.
\end{displaymath}
Note that,  as concerned to these indices, we are working with
two-component spinors and adopt Penrose's conventions of the
book~\cite{Stewart1996}. We follow those conventions even when
complex conjugation of spinor and pseudo-conjugation of Grassmann
quantities are involved.

In the adjoint representation~\cite{Ilyenko2003} the matrices
$T_a$ and $\tau_A$ can be written as follows (solid lines are
drawn to emphasize their block structure):
\begin{widetext}
\begin{displaymath}
 T_1 = \left( \mbox{%
\begin{tabular}{ccc|cc}
 $0$&$0$&$i$& $0$&$0$ \\
 $0$&$0$&$0$& $0$&$0$ \\
$-i$&$0$&$0$& $0$&$0$ \\ \hline
 $0$&$0$&$0$& $0$&$1/2$ \\
 $0$&$0$&$0$& $1/2$&$0$ \\
\end{tabular} }
              \right),\;\;\;
T_2 = \left(\mbox{%
\begin{tabular}{ccc|cc}
$0$&$-i$&$0$& $0$&$0$ \\
 $i$&$0$&$0$& $0$&$0$ \\
 $0$&$0$&$0$& $0$&$0$ \\ \hline
 $0$&$0$&$0$& $0$&$-i/2$ \\
 $0$&$0$&$0$& $i/2$&$0$ \\
\end{tabular} }
              \right),\;\;\;
T_3 = \left(\mbox{%
\begin{tabular}{ccc|cc}
 $0$&$0$&$0$& $0$&$0$ \\
$0$&$0$&$-i$& $0$&$0$ \\
 $0$&$i$&$0$& $0$&$0$ \\ \hline
 $0$&$0$&$0$& $1/2$&$0$ \\
 $0$&$0$&$0$& $0$&$-1/2$ \\
\end{tabular} }
             \right),
\end{displaymath}
\begin{equation}
 T_4 \equiv \tau_1 = \frac{1}{2}\left(\mbox{%
\begin{tabular}{ccc|cc}
 $0$&$0$&$0$& $0$&$1$ \\
 $0$&$0$&$0$& $-1$&$0$ \\
 $0$&$0$&$0$& $-i$&$0$ \\ \hline
$-i$&$0$&$0$& $0$&$0$ \\
$0$&$-i$&$1$& $0$&$0$ \\
\end{tabular} }
              \right),\;\;\;
  T_5 \equiv \tau_2 = \frac{1}{2}\left(\mbox{%
\begin{tabular}{ccc|cc}
  $0$&$0$&$0$& $1$&$0$ \\
  $0$&$0$&$0$& $0$&$1$ \\
  $0$&$0$&$0$& $0$&$-i$ \\ \hline
$0$&$-i$&$-1$& $0$&$0$ \\
  $i$&$0$&$0$& $0$&$0$ \\
\end{tabular} }
              \right).
\label{AdjRepM}
\end{equation}
\end{widetext}
Let us denote $T_4$ = $\tau_1$, $T_5$ = $\tau_2$ and employ
lowercase Greek indices from the beginning of the alphabet
($\alpha$, $\beta$, 
etc.) to run over the whole set, $T_\alpha$, of the generators of
\textit{osp}(2/1;~$\mathbb{C}$). We then find~\cite{Ilyenko2003}
that the non-degenerate super-Killing form, $B(T_\alpha,
T_\beta)$, is given by
\begin{equation}
B(T_\alpha, T_\beta) = \frac{2}{3}\,\mbox{str}(T_\alpha T_\beta) =
\left(\mbox{%
\begin{tabular}{c|c}
$\delta_{ab}$& $0$ \\ \hline
          $0$& $i\epsilon_{AB}$ \\
\end{tabular}}
 \right),
\label{sKillingF}
\end{equation}
where the supertrace operation is adopted from
\cite[pp.~18-19,~42]{Cornwell1989}.

It turns out that not all of the \textit{osp}(1$|$2;~$\mathbb{C}$)
algebra generators are hermitian. A proper generalization of the
hermitian conjugation~\cite{RS1978} (\textit{graded}
\textit{adjoint}) is denoted by (${}^\ddag$): on the even
generators the operation coincides with ordinary hermitian
conjugation (${}^+$) while the odd ones obey more complicated
relations (see below and at the end of Sec.~\ref{RepSp}).
Following the paper~\cite{SNR19771}, we call them the grade star
hermiticity conditions:
\begin{equation}\label{gsHC}
\tau_{\pm}^\ddagger = \pm\tau_{\mp},
\end{equation}
where we denoted $\tau_{\pm}$ = $\tau_1$ $\pm$ $i\tau_2$.

Let us consider complex-valued matrices divided into blocks
according to the scheme (cf. (\ref{AdjRepM}) and
(\ref{sKillingF})):
\begin{eqnarray}
M_{even} = \left(\mbox{{\small%
\begin{tabular}{c|c}
$A$& $0$ \\ \hline
$0$& $D$ \\
\end{tabular}}} \right) &
\mbox{and} &
M_{odd} = \left(\mbox{{\small%
\begin{tabular}{c|c}
$0$& $B$ \\ \hline
$C$& $0$ \\
\end{tabular}}} \right),
\end{eqnarray}
where, for the purposes of this paper, $B$ and $C$ are 2$\times$3
rectangular blocks and $A$ and $D$ are 3$\times$3 and 2$\times$2
square blocks, respectively. On these matrices the supertrace
operation gives str$M_{even}$~= tr$A$ $-$ tr$D$ and str$M_{odd}$~=
0 (here ``tr'' denotes the ordinary trace) while the grade star
hermiticity condition reads
\begin{eqnarray}
M_{even}^\ddag = \left(\mbox{{\small%
\begin{tabular}{c|c}
$A^+$& $0$ \\ \hline
$0$& $D^+$ \\
\end{tabular}}} \right) &
\mbox{and} &
M_{odd}^\ddag = \left(\mbox{{\small%
\begin{tabular}{c|c}
$0$& $-C^+$ \\ \hline
$B^+$& $0$  \\
\end{tabular}}} \right).
\end{eqnarray}

We shall also use multiplication of algebra generators by scalars.
Such an operation must take into account that Grassmann-odd
scalars anticommute with the odd algebra generators while commute
with complex numbers and the even algebra
generators~\cite{GK1985}. The following construction possesses all
of these properties. Let $a$ be a scalar and deg\,$a$ be its
degree (0 or 1 depending on whether it is Grassmann-even or
Grassmann-odd, respectively). Then multiplication by $a$ is
defined as follows:
\begin{widetext}
\begin{eqnarray}
aM_{odd}\hspace{0.5em} =
\left(\mbox{{\small%
\begin{tabular}{c|c}
$a$& $0$ \\ \hline
$0$& $(-1)^{\mbox{deg\,\textit{a}}}a$ \\
\end{tabular}}} \right)
\left(\mbox{{\small%
\begin{tabular}{c|c}
$0$& $B$ \\ \hline
$C$& $0$ \\
\end{tabular}}} \right)%
& = & (-1)^{\mbox{deg\,\textit{a}}}
\left(\mbox{{\small%
\begin{tabular}{c|c}
$0$& $B$ \\ \hline
$C$& $0$ \\
\end{tabular}}} \right)
\left(\mbox{{\small%
\begin{tabular}{c|c}
$a$& $0$ \\ \hline
$0$& $(-1)^{\mbox{deg\,\textit{a}}}a$ \\
\end{tabular}}} \right)
=  \nonumber \\
& & \hphantom{\left(\mbox{{\small%
\begin{tabular}{c|c}
$A$& $0$ \\ \hline
$0$& $D$ \\
\end{tabular}}} \right)
\left(\mbox{{\small%
\begin{tabular}{c|c}
$a$& $0$ \\ \hline
$0$& $(-1)^{\mbox{deg\,\textit{a}}}a$ \\
\end{tabular}}} \right)}
= (-1)^{\mbox{deg\,\textit{a}}}M_{odd}a,  \nonumber \\
 aM_{even} =
\left(\mbox{{\small%
\begin{tabular}{c|c}
$a$& $0$ \\ \hline
$0$& $(-1)^{\mbox{deg\,\textit{a}}}a$ \\
\end{tabular}}} \right)
\left(\mbox{{\small%
\begin{tabular}{c|c}
$A$& $0$ \\ \hline
$0$& $D$ \\
\end{tabular}}} \right)%
& = &
\left(\mbox{{\small%
\begin{tabular}{c|c}
$A$& $0$ \\ \hline
$0$& $D$ \\
\end{tabular}}} \right)
\left(\mbox{{\small%
\begin{tabular}{c|c}
$a$& $0$ \\ \hline
$0$& $(-1)^{\mbox{deg\,\textit{a}}}a$ \\
\end{tabular}}} \right)
=M_{even}a. \nonumber
\end{eqnarray}%
\end{widetext}

\section{The group property}
\label{BCH-f}%

Given a Lie algebra one can turn over to a Lie group by
exponentiating the generators multiplied by transformation
parameters. This, in a usual fashion, gives us the gauge
transformations. In the case of a graded Lie algebra we are faced
with a problem: anticommutators seem to rule out the application
of the Baker-Campbell-Hausdorff formula, which is necessary to
prove that subsequent transformations do not leave the group
manifold. This problem is solved via introduction of Grassmann-odd
parameters (cf.~\cite{Berezin1970}). In the case under
consideration these are Grassmann-odd \textit{su}(2)-spinors
$\xi^A$, $\theta^A$, etc., which multiply the odd generators. They
are included on equal footing with ordinary (Grassmann-even)
parameters $\varepsilon^a$ multiplying the even generators
(hopefully, there will not be confusion about use the same kernel
letter, $\varepsilon$, to denote a Grassmann-even transformation
parameter and the Levi-Civita totally antisymmetric symbol in
three dimensions). By definition, $\xi^A$, $\theta^A$, etc.
satisfy
\begin{equation*}
[\varepsilon^a,\,\theta^A]=0,\,\,\{\xi^A,\,\xi^B\}=
\{\theta^A,\,\theta^B\}=0,\,\,\{\xi^A,\,\theta^B\}=0.
\end{equation*}
Then, the necessary relations can be given in terms of commutators
only:
\begin{equation*}
[\xi^A\tau_A,\,\theta^B\tau_B]=-\frac{i}{2}(\xi^{\{A}\theta^{B\}}
+ \xi^{[A}\theta^{B]})(\sigma^a)_{AB}T_a,%
\end{equation*}
where $\xi^{\{A}\theta^{B\}} = 1/2(\xi^A\theta^B + \xi^B\theta^A)$
and $\xi^{[A}\theta^{B]} = 1/2(\xi^A\theta^B - \xi^B\theta^A)$ are
convenient shorthand notations. This result was obtained using
anticommutator for odd generators in definition (\ref{DefRel}).
Using a fundamental fact of spinor algebra,
$\epsilon_{AB}\epsilon_{CD}$ $+$ $\epsilon_{AC}\epsilon_{DB}$ $+$
$\epsilon_{AD}\epsilon_{BC}$ $=$ $0$, one can calculate
\begin{equation*}
\xi^{[A}\theta^{B]}=\frac{1}{2}(\xi_C\theta^C)\epsilon^{AB}.
\end{equation*}
From symmetry of $(\sigma^a)_{AB}$ in the uppercase indices, it
then follows that
\begin{equation}\label{ComOdd}
[\xi^A\tau_A,\,\theta^B\tau_B] =
-\frac{i}{2}\xi^{\{A}\theta^{B\}}(\sigma^a)_{AB}T_a \equiv
-\frac{i}{2}[\xi^{A},\theta^{B}](\sigma^a)_{AB}T_a
\end{equation}
and, in particular, the commutator
$[\theta^A\tau_A,\,\theta^B\tau_B]$ vanishes identically. One can
also calculate
\begin{equation}
\left[\kappa^aT_a,\,\varepsilon^bT_b\right] =
i\kappa^{[a}\varepsilon^{b]}\varepsilon_{abc}T_c\,\,\,\,
\mbox{and}\,\,\,\,
\left[\varepsilon^{a}T_{a},\,\theta^{A}\tau_{A}\right] =
\tilde{\theta}{}^{B}\tau_{B}, \label{ComEve}
\end{equation}
where $2\tilde{\theta}{}^B$ $=$
$\varepsilon^a$$\theta^A$$(\sigma_a)_A^{\hphantom{A}B}$ is again a
Grassmann-odd transformation parameter.

Group elements of UOSp(1$|$2) are obtained by exponentiating the
algebra
\begin{equation}
U(\varepsilon,\,\theta) = \exp(i(\varepsilon^aT_a + \theta^A\tau_A))%
\label{SPT}
\end{equation}
and the Baker-Campbell-Hausdorff formula,
\begin{equation}
\exp(M)\exp(N) = \exp(M + N + \frac{1}{2}[M,\,N] + \hdots),
\label{BCH_f}
\end{equation}
may be applied to determine motion in the parameter space under a
(left) multiplication with a group element $U(\kappa,\,\xi)$:
\begin{equation*}
U(\varepsilon' ,\,\theta') =
U(\kappa,\,\xi)U(\varepsilon,\,\theta).
\end{equation*}

\section{Infinitesimal transformations and reality conditions}

Let us examine expression (\ref{BCH_f}) restricting ourselves by
taking into account the first non-trivial contribution, i.e. the
two-fold commutator $[M, N]$. Writing $M$ = $i(\varepsilon^aT_a$ +
$\theta^A\tau_A)$ and $N$ = $i(\kappa^aT_a$ + $\xi^A\tau_A)$, we
have
\begin{equation*}
i(\varepsilon'{}^aT_a + \theta'{}^A\tau_A) = M + N +
\frac{1}{2}[M,\,N] + \ldots,
\end{equation*}
where dots denote the sum of linear combinations of $k$-fold ($k
> 2$, $k \in \mathbb{N}$) commutators of $M$ and $N$~\cite{MKS1966}.
Substituting expressions for $M$, $N$ and using (\ref{ComOdd}), we
obtain after some algebra
\begin{eqnarray}
\varepsilon'{}^a & = & \varepsilon^a + \kappa^a -
\frac{1}{2}\kappa_b\varepsilon_c\,\varepsilon^{bca} +
\frac{1}{4}[\xi^{A},\theta^{B}](\sigma^a)_{AB} + \ldots, \nonumber \\
\theta'{}^A & = & \theta^A + \xi^A + \frac{i}{4}(\kappa_b\theta^B
- \varepsilon_b\,\xi^B)(\sigma^b)_{B}^{\hphantom{B}A} + \ldots%
\label{PTrans}
\end{eqnarray}
Here again dots denote the contribution from the sum of linear
combinations of $k$-fold ($k > 2$, $k \in \mathbb{N}$)
commutators. The first three summands in the first row of formula
(\ref{PTrans}) reflect the non-commutative character of the proper
Lie subalgebra, \textit{su}(2), of
\textit{osp}(1$|$2;~$\mathbb{C}$) the last one being contribution
from the odd part of the graded Lie algebra. The last summand in
the second row of the formula is obviously a Grassmann-odd
quantity, and it reflects the non-commutative property of the even
and odd parts of the graded Lie algebra.

In the view of intended applications, contribution from
Grassmann-odd part of the algebra into the law of composition of
Grassmann-even parameters needs to be investigated in more detail.
First, let us calculate that
\begin{equation}
  2[\xi^A,\theta^B](\sigma^a)_{AB} =
  \xi_A\epsilon^{AB}(\sigma^a)_{B}^{\hphantom{B}C}\theta_C -
  \theta_A\epsilon^{AB}(\sigma^a)_{B}^{\hphantom{B}C}\xi_C
  = \xi^t\Sigma\sigma^a\theta - \theta^t\Sigma\sigma^a\xi,
\label{CVect}
\end{equation}
where we employed some self-evident matrix notations; the
superscript (${}^t$) denotes transpose. Comparing the result
(\ref{CVect}) and a description of \textit{su}(2)-spinors of
3D~Euclidean space in the book \cite[p.~48]{Cartan1966}, one
immediately realizes that the last term of the first equation in
system (\ref{PTrans}) is, in general, a \textit{complex} vector of
3D~Euclidean space. Second, the representation (\ref{CVect}) tells
us that components of this vector vanish if $\xi_A$=$\theta_A$ as
required by a property of a one-parameter subgroup of
transformations (\ref{SPT}). Finally, this vector also has all
components equal to zero if $\xi_A$ = $-\,\theta_A$. This shows
that the inverse of the group element $U(\varepsilon,\,\theta)$
has the form
\begin{equation}
U^{-1}(\varepsilon,\,\theta) = \exp[-i(\varepsilon^aT_a + \theta^A\tau_A)].%
\label{ISPT}
\end{equation}

If one intends, as customarily done in a meaningful Yang-Mills
theory, to treat $\varepsilon^a$, $\kappa^a$, etc. as real-valued
Grassmann-even transformation parameters, then it is necessary to
impose some conditions on the \textit{su}(2)-spinors $\xi_A$,
$\theta_A$, etc. in order to ensure that (\ref{CVect}) will be a
\textit{real} 3D~Euclidean vector. Such a condition must be
compatible with transformation properties of the corresponding
space of \textit{su}(2)-spinors, $\xi_A$, and take into account
that its members are also Grassmann-odd quantities. In fact, this
condition should involve a passage from an \textit{su}(2)-spinor
to its conjugate and, thus, rely on the definition of an
anti-involution in the space of spinors (see, e.g.
\cite[p.~100]{Cartan1966}). Let us observe first that for a
Grassmann algebra on one generator the last term in the first
relation in (\ref{PTrans}) vanishes identically. This is a
somewhat trivial situation. The next non-trivial one arises when
all \textit{su}(2)-spinors under consideration take values in a
Grassmann algebra on two odd generators (more generally on even or
infinite number of odd generators, cf.~\cite{SW2005}), $\beta_1$
and $\beta_2$: $\beta_1^2$ = $\beta_2^2$ = 0, $\beta_1\beta_2$ =
$-\beta_2$$\beta_1$ (see, e.g.~\cite[p.~7]{Cornwell1989}). We
employ lowercase Roman indices from the middle of the alphabet
running over 1 and 2 to enumerate the decompositions of various
quantities in the corresponding basis of the Grassmann algebra.
Decomposing $\xi_A$ and $\theta_A$ into this basis one obtains
\begin{displaymath}
\xi_A = \xii{}_A\beta_i\,\,\,\mbox{and} \,\,\,\theta_B =
\thetaj{}_B\beta_j,
\end{displaymath}
where $\xii{}_A$ and $\thetaj{}_B$ are ordinary, i.e.
complex-valued Grassmann-even, \textit{su}(2)-spinors of
3D~Eclidean space, and summation over repeated indices is assumed.
In this case we can write
\begin{equation}
[\xi^A,\theta^B](\sigma^a)_{AB} =
2\beta_1\beta_2(\xione^t\Sigma\sigma^a\thetatwo -
\thetaone^t\Sigma\sigma^a\xitwo).%
\label{CVectG}
\end{equation}
Now we impose some additional conditions on \textit{su}(2)-spinors
$\xii{}_A$, $\thetaj{}_A$, etc. to ensure that (\ref{CVectG})
gives a \textit{real} Grassmann-even 3D~Euclidean vector. One way
of doing so in a manner preserving all the spinor transformations
properties (Majorana conditions) is to define~\cite{Ilyenko2003}
\begin{equation}
\xione{}_A = iC_{A}^{\hphantom{A}B'}\xibartwo{}_{B'}, \,\,\,
\thetaone{}_A = iC_{A}^{\hphantom{A}B'}\thetabartwo{}_{B'}, \,\,\,
\mbox{etc.},%
\label{CC}
\end{equation}
where the `charge conjugation' matrix $C$ ($C\overline{C}$ =
${}-I$) is given by
\begin{displaymath}
    C = \|C_{A}^{\hphantom{A}B'}\| =
\left( \begin{array}{cc}
      0 & 1 \\
\!\! -1 & 0
\end{array} \right)
= \|\overline{C}_{A'}^{\hphantom{A'}B}\| = \overline{C}.
\end{displaymath}
In (\ref{CC}) a bar over the spinors in the left-hand sides of the
relations and primes over the indices denote complex conjugation
($\xibari{}_{A'}$~= $(\xii{}_A)^*$). The `charge conjugation'
matrix, $C_A^{\hphantom{A}B'}$, is responsible for invariant
preservation of spinor properties (for details see, e.g., the
review article~\cite[pp.~108~--~109]{Rashevskii1955}, where this
object is denoted by $\stackrel{*}{\Pi}{}^\lambda_{\dot{\mu}}$;
also compare with treatment in \cite{Cartan1966}). Note that
definitions (\ref{CC}) are essentially \textit{the proper
generalization of reality conditions} from complex numbers to
spinors. As also seen from that definitions, each Grassmann-odd
\textit{su}(2)-spinor $\xi_A$, $\theta_A$, etc. is defined by a
single ordinary \textit{su}(2)-spinor. For the sake of notations
denoting
\begin{displaymath}
\eta_A = \xitwo{}_A\,\,\,\mbox{and}\,\,\,\vartheta_B =
\thetatwo{}_B,
\end{displaymath}
respectively, we write
\begin{equation}\label{RVect}
v^a \equiv \xione^t\Sigma\sigma^a\thetatwo -
\thetaone^t\Sigma\sigma^a\xitwo =
i(\bar{\eta}{}^tC{}^t\Sigma\sigma^a\vartheta -
\bar{\vartheta}{}^tC{}^t\Sigma\sigma^a\eta).
\end{equation}
On comparison with \cite{Cartan1966}, one can check that
$\varkappa^a$ is indeed a \textit{real} 3D~Euclidean
vector~\cite{Ilyenko2003}. In components
it reads:%
\begin{eqnarray}\label{RVectComps}
v^1 & = & i(\bar{\eta}{}_{1'}\vartheta_2 -
\bar{\vartheta}{}_{2'}\eta_1 + \bar{\eta}{}_{2'}\vartheta_1 -
\bar{\vartheta}{}_{1'}\eta_2), \nonumber \\
v^2 & = & \hphantom{i(}\bar{\eta}{}_{2'}\vartheta_1 +
  \bar{\vartheta}{}_{1'}\eta_2 - \bar{\eta}{}_{1'}\vartheta_2 -
  \bar{\vartheta}{}_{2'}\eta_1, \\
v^3 & = & i(\bar{\eta}{}_{1'}\vartheta_1 -
\bar{\vartheta}{}_{1'}\eta_1
  - \bar{\eta}{}_{2'}\vartheta_2 + \bar{\vartheta}{}_{2'}\eta_2). \nonumber
\end{eqnarray}
These are obviously real quantities and the vector $v^a$ vanishes
if and only if $\eta_A$ = $\pm\vartheta_A$ as required.

\section{Action on a representation space}
\label{RepSp}

Having formulated meaningful reality conditions, we are in
position to explore action of the group element (\ref{SPT}) on a
suitable representation vector space.

First, let us observe that because of definition of the matrix $U$
by its Taylor's expansion, the fact that the generators $T_a$ and
$\tau_A$ are block and off-block diagonal, respectively, their
multiplication properties and those of the Grassmann-even and
Grassmann-odd transformation parameters, it is easy to see that
any matrix $U$ has a specific decomposition
\begin{equation}\label{GrEl}
U = \left(\mbox{{\small%
\begin{tabular}{c|c}
$A$& $B$ \\ \hline
$C$& $D$ \\
\end{tabular}}} \right),
\end{equation}
where $A$ is a ($r \times p$) sub-matrix, $B$ is a ($s \times p$)
sub-matrix, $C$ is a ($r \times q$) sub-matrix and $D$ is a ($s
\times q$) sub-matrix. Following nomenclature of the book
\cite{Cornwell1989}, we call the matrix $U$ a ($p/q\times r/s$)
super-matrix. Moreover, the sub-matrices $A$ and $D$ have
contributions only from an even number of $\tau$s' multipliers
and, hence, only even multipliers of Grassmann-odd transformation
parameters $\theta$s' are present there. Thus, elements of those
sub-matrices are in the even subspace, $\mathbb{C}B_{L0}$, of the
complex Grassmann algebra (for more details see the book
\cite[pp.~10--11]{Cornwell1989}). The sub-matrices $B$ and $C$ by
an analogues argument include an odd number of $\tau$s' and
$\theta$s' multipliers and, hence, are in the odd subspace,
$\mathbb{C}B_{L1}$, of the complex Grassmann algebra. Therefore,
any such a super-matrix $U$ is an \textit{even} super-matrix and
by the results of the previous section such matrices form a
supergroup. Furthermore, by constraction any such a super-matrix
is invertible.

Second, consider \textit{even} super-column $\Psi$ (($p/q\times
0/1$) super-matrices) and super-row $\Phi$ (($1/0\times r/s$)
super-matrices) vectors:
\begin{equation*}
\Psi = \left(\mbox{{\small%
\begin{tabular}{c}
$\Psi_1$\\ \hline
$\Psi_2$\\
\end{tabular}}} \right)%
\,\,\,\,\mbox{and}\,\,\,\,
\Phi = \left(\mbox{{\small%
\begin{tabular}{c|c}
$\Phi_1$&$\Phi_2$\\
\end{tabular}}} \right),
\end{equation*}
where $\Psi_1$ and $\Phi_1$ are ($1 \times p$) and ($r \times 1$)
sub-matrices, $\Psi_2$ and $\Phi_2$ are ($1 \times q$) and ($s
\times 1$) sub-matrices, respectively. The elements of $\Psi_1$
and $\Phi_1$ are Grassmann-even and those of $\Psi_2$ and $\Phi_2$
are Grassmann-odd entities. \textit{Action of even super-matrices}
$U$ \textit{on such even super-column}(\textit{row})
\textit{vectors transform them again into even
super-column}(\textit{row}) \textit{vectors}.

Third, since the sub-matrices $B$ and $C$ are in the odd sub-space
of $\mathbb{C}B_{L1}$, one needs to modify the very notion of
complex conjugation. Instead, the operation called
\textit{pseudo}-\textit{conjugation} is to be
used~\cite{RS1978,SW2005}, which is one of at least two
inequivalent generalizations of complex conjugation to
supernumbers (cf.~\cite{DeWitt1992}); it coincides with ordinary
complex conjugation (denoted above by the asterisk (${}^*$)) on
Grassmann-even quantities (e.g. ordinary complex numbers) being
there an involution and is an anti-involution on the Grassmann-odd
ones. Following~\cite{SW2005}, it will be denoted by a superscript
diamond (${}^{\diamond}$).

For the sake of argument let $U$ be ($1/1 \times 1/1$) matrices
(see (\ref{GrEl})), the actual size can be easily treated the same
way, and let also $\Psi$ be a ($1/1 \times 0/1$) even super-column
vector as regarded to the linear transformations defined below.
The entries $\Psi_1$ and $\Psi_2$ themselves could be, for
example, Dirac bispinors. Consider a linear transformation
\begin{equation}\label{LinTr}
\left(\mbox{{\small%
\begin{tabular}{c}
$\Psi^\prime_1$\\ \hline
$\Psi^\prime_2$\\
\end{tabular}}} \right)%
=
\left(\mbox{{\small%
\begin{tabular}{c}
$A\Psi_1 + B\Psi_2$\\ \hline
$C\Psi_1 + D\Psi_2$\\
\end{tabular}}} \right)%
\equiv U\Psi.
\end{equation}
Taking transposition of each line in (\ref{LinTr}) (it acts on
$\Psi$'s) and pseudo-conjugate as well as introducing the
(modified to supernumbers) Dirac conjugation by
$\bar{\Psi}^\prime_i$ = $(\Psi^t_i)^{\diamond}\gamma_0$, we
obtain:
\begin{eqnarray}
\left(\mbox{{\small%
\begin{tabular}{c|c}
$\bar{\Psi}^\prime_1$&$\bar{\Psi}^\prime_2$\\
\end{tabular}}} \right)%
& = &
\left(\mbox{{\small%
\begin{tabular}{c|c}
$\bar{\Psi}_1 A^{\diamond} - \bar{\Psi}_2
B^{\diamond}$&$\bar{\Psi}_1 C^{\diamond} +
\bar{\Psi}_2 D^{\diamond}$\\
\end{tabular}}} \right) \nonumber \\%
& = &
\left(\mbox{{\small%
\begin{tabular}{c|c}
$\bar{\Psi}_1$&$\bar{\Psi}_2$\\
\end{tabular}}} \right)%
\left(\mbox{{\small%
\begin{tabular}{c|c}
$A^{\diamond}$& $C^{\diamond}$ \\ \hline
$-B^{\diamond}$& $D^{\diamond}$ \\
\end{tabular}}} \right),
\end{eqnarray}
where the Grassmann character of the involved quantities has been
taken into account. Recall that for any super-matrix $U$
partitioned as in (\ref{GrEl}) the super-transpose is defined by
\begin{equation*}
U^{st} = \left(\mbox{{\small%
\begin{tabular}{c|c}
$A^t$& $(-1)^{\mbox{deg\,\textit{U}}}C^t$ \\ \hline
$-(-1)^{\mbox{deg\,\textit{U}}}B^t$& $D^t$ \\
\end{tabular}}} \right),
\end{equation*}
where $({}^t)$ denotes the ordinary transposition; for
\textit{even} super-column(row) vectors this implies:
\begin{equation*}
\Psi^{st} = \left(\mbox{{\small%
\begin{tabular}{c|c}
$\Psi^{t}_1$&$\Psi^t_2$\\
\end{tabular}}} \right)%
\,\,\,\,\mbox{and}\,\,\,\,
\Phi^{st} = \left(\mbox{{\small%
\begin{tabular}{c}
$\hphantom{-}\Phi^t_1$\\
$-\Phi^t_2$\\
\end{tabular}}} \right).
\end{equation*}
It then follows that
\begin{equation*}
\left(\mbox{{\small%
\begin{tabular}{c|c}
$A^{\diamond}$& $C^{\diamond}$ \\ \hline
$-B^{\diamond}$& $D^{\diamond}$ \\
\end{tabular}}} \right)%
=
\left(\mbox{{\small%
\begin{tabular}{c|c}
$A$& $C$ \\ \hline
$-B$& $D$ \\
\end{tabular}}} \right)^{\diamond}%
=
\left(\left(\mbox{{\small%
\begin{tabular}{c|c}
$A$& $B$ \\ \hline
$C$& $D$ \\
\end{tabular}}} \right)^{st}\right)^{\diamond},
\end{equation*}
i.e. if as in (\ref{LinTr})
\begin{equation}
\Psi^\prime = U\Psi \,\,\,\, \mbox{then} \,\,\,\,
\bar{\Psi}^\prime = \bar{\Psi}U^{\ddag},
\end{equation}
thus, generalizing the corresponding result in the Yang-Mills
theory. As one can easily check, the {\it pseudo-conjugate} {\it
super-transpose} possesses all the properties of the
\textit{graded} \textit{adjoint} (${}^\ddag$)~\cite{RS1978}.

\section{Proof of graded unitary property}

One needs to verify that $U^{\ddag} = U^{-1}$ with the given
definition of (${}^\ddag$). If $U = \exp{(M)}$ then the
Backer-Campbell-Hausdorff formula implies that one should have
$M^{\ddag} = -M$, where $M = i(\varepsilon^aT_a +
\theta^A\tau_A)$.

Considering the example with a Grassmann algebra on two odd
generators (generalization to Grassmann algebras with any even or
infinite number of odd generators is straightforward), we
have~\cite{SW2005}
\begin{equation}\label{PC}
\beta^{\diamond}_{1} = -\beta_{2},\,\,\,\, \beta^{\diamond}_{2} =
\beta_{1}.
\end{equation}
Recall that for $\theta_A = \thetaj{}_A\beta_{j}$ we defined
in~\eqref{CC}
\begin{eqnarray*}
\thetaone{}_A & = &\,\,\,\,
iC_{A}^{\hphantom{A}B'}\thetabartwo{}_{B'},\,\,
\thetatwo{}_A = -iC_{A}^{\hphantom{A}B'}\thetabarone{}_{B'}, \\
\thetabarone{}_{A'}\! & = &
-i\overline{C}_{A'}^{\hphantom{A}B}\thetatwo{}_{B},\,\,\,\,
\thetabartwo{}_{A'} =\,\,\,\,
i\overline{C}_{A'}^{\hphantom{A}B}\thetaone{}_{B},
\end{eqnarray*}
and that on \textit{osp}(1$|$2;~$\mathbb{C}$) generators $T_{a}$
and $\tau_{A}$ the pseudo-conjugation coincides with complex
conjugation. A direct calculation shows that
\begin{equation}\label{GrConGen}
(T_{a})^{\ddag} \equiv (T^{\ddag})_{a} = T_{a},\,\, \mbox{and}
\,\, (\tau_{A})^{\ddag} \equiv (\tau^{\ddag})_{A'} =
-i\overline{C}_{A'}^{\hphantom{A}B}\tau_{B}.
\end{equation}
The former equation in~\eqref{GrConGen} is just a restatement of
hermiticity of the even generators while the later once again
exhibits a strong connection between Grassmann-odd sector of the
compact graded Lie algebra \textit{osp}(1$|$2;~$\mathbb{C}$) on
one side and 3D~Euclidean spinors on the other (cf.~\eqref{gsHC}).
It then follows that
\begin{eqnarray}\label{sub1}
M^\ddag & = & ((i(\varepsilon^aT_a +
\theta^A\tau_A)){}^{st}){}^{\diamond} =
-i(\varepsilon^a(T_a)^{\ddag} +
(\theta^A)^{\diamond}(\tau_A)^{\ddag}) = \nonumber \\
& = & -i\varepsilon^aT_a - (\thetaj{}^A\beta_{j}){}^{\diamond}
\overline{C}_{A'}^{\hphantom{A}B}\tau_{B} = -i\varepsilon^aT_a -
(\bar{\thetaj}{}^{A'}\beta^{\diamond}_{j})
\overline{C}_{A'}^{\hphantom{A}B}\tau_{B}
\end{eqnarray}
(it is easy to see that for a generic $\varepsilon^a =
\tilde{\varepsilon}{}^a + \tilde{e}{}^a\beta_1\beta_2$ with
coefficients $\tilde{\varepsilon}{}^a$ and $\tilde{e}{}^a$ real in
the ordinary sence the property $(\varepsilon{}^a){}^{\diamond} =
\varepsilon{}^a$ holds because of~\eqref{PC} and the Grassmann-odd
nature of generators $\beta_j$). Here we expand with the use
of~\eqref{PC}
\begin{eqnarray}\label{sub2}
(\bar{\thetaj}{}^{A'}\beta^{\diamond}_{j})\overline{C}_{A'}^{\hphantom{A}B}
& = & (\epsilon^{A'D'}[\thetabarone{}_{D'}\beta^{\diamond}_{1} +
\thetabartwo{}_{D'}\beta^{\diamond}_{2}])
\overline{C}_{A'}^{\hphantom{A}B} =
i\epsilon^{A'D'}\overline{C}_{D'}^{\hphantom{D}C}(\thetatwo{}_{C}\beta_{2}
+ \thetaone{}_{C}\beta_{1})\overline{C}_{A'}^{\hphantom{A}B} = \nonumber \\
& = &
i\overline{C}_{A'}^{\hphantom{A}B}\epsilon^{A'D'}\overline{C}_{D'}^{\hphantom{D}C}
(\thetaj{}_{C}\beta_{j}) =
i\overline{C}_{A'}^{\hphantom{A}B}\epsilon^{A'D'}\overline{C}_{D'}^{\hphantom{D}C}
\theta_{C},
\end{eqnarray}
where in the first summand in the parenthesis in the last equality
of the first line an important cancelation of minus signs happens
both because of definition of \textit{reality condition} on
spinors~\cite{Ilyenko2003} (see also~\cite{Cartan1966,KW2006}) and
\textit{pseudo}-\textit{conjugation} on Grassmann numbers adopted
from~\cite{SW2005}. Using the identity
$\overline{C}\,{}^{t}\overline{\Sigma}\,\overline{C} = \Sigma$ or
via direct calculation, one can also check that the relation
\begin{equation}\label{sub3}
\overline{C}_{A'}^{\hphantom{A}B}\epsilon^{A'D'}\overline{C}_{D'}^{\hphantom{D}C}
= \epsilon^{BC}
\end{equation}
holds. From~\eqref{sub1} with the aid of~\eqref{sub2}
and~\eqref{sub3}, we, finally, obtain
\begin{equation*}
M^\ddag = -i\varepsilon^aT_a - i\epsilon^{BC}\theta_{C}\tau_{B}
\equiv -i(\varepsilon^aT_a + \theta^{A}\tau_{A}) = -M,
\end{equation*}
thus showing that the (Grassmann-valued) matrix $M$ is
\textit{graded} \textit{anti}-\textit{hermitian}. It then
immediately follows that the group element~\eqref{SPT} of
UOSp(1$|$2) is \textit{graded} \textit{unitary}
\begin{equation*}
U^\ddag = U^{-1}
\end{equation*}
proving that the corresponding group is a graded unitary
orthosymplectic group.

\section{Conclusions and Outlook}

We have accomplished algebraic preliminaries necessary to check
the gauge invariance of the proposed field
strength~\cite{BL1996,Ilyenko2001} for UOSp(1$|$2) graded
Yang-Mills theory on 4D~Minkowski space-time and to develop an
analogue of `non-commutative electrodynamics' with massive matter
fields. These required utilization of 3D~Euclidean spinors,
reality conditions on them~\cite{Ilyenko2003} (see
also~\cite{Cartan1966,KW2006}), and notion of pseudo-conjugation
on Grassmann-valued quantities~\cite{SW2005,RS1978}.

In the conventional Yang-Mills theory the number of generators of
the underlying Lie algebra correspond to the number of gauge
bosons. In this respect we shall be interested in exploring the
role of Grassmann-odd generators of
\textit{osp}(1$|$2;~$\mathbb{C}$) in such a graded Yang-Mills
theory. It will be also interesting to investigate the properties
of the Grassmann-odd sector of the representation space for such a
graded generalization of `non-commutative electrodynamics' with
massive matter content. Another important question is whether
there exists an analogues connection between Euclidean spinors and
other compact graded Lie algebras (cf.~\cite{Cornwell1989}).

\section*{Acknowledgements}

I would like to thank Prof.~M.~Scheunert for pointing out to me
the articles~\cite{SW2005,KW2006} and Prof.~Yu.P.~Stepanovsky,
Drs.~V.~Gorkavyi, V.~Pidstrigach, and C.J.~Wainwright for helpful
discussions. I am also grateful to Dr.~T.S.~Tsou for interest to
this work.



\begin{thebibliography}{99}
\bibitem{Berezin1970} F.A.~Be\-re\-zin and G.I.~Kac, {\it Lie groups
  with commuting and anticommuting parameters}, Math. USSR~--~Sb.
  {\bf 11} (1971) 311.
\bibitem{CNS1975} L.~Corwin, Y.~Ne'eman and S.~Stenberg,
  {\it Graded Lie algebras in mathematics and physics (Bose-Fermi
  symmetry)}, Rev. Mod. Phys. {\bf 47} (1975) 573.
\bibitem{Kac1977} V.~Kac, {\it Representations of classical Lie
  superalgebras}, LMS Lecture Notes {\bf 676} (1977) 597.
\bibitem{Neeman1979} Y.~Ne'eman, {\it Irreducible gauge theory of a
  consolidated Salam-Weinberg model}, Phys.~Lett. B {\bf 81} (1979) 190.
\bibitem{BL1996} R.~Brooks and A.~Lue, {\it The monopole equations
  in topological Yang-Mills}, J. Math. Phys. {\bf 37} (1996) 1100.
\bibitem{Ilyenko2001} K.~Ilyenko, {\it Field strength for graded Yang-Mills
  theory}, Problems Atom. Sci. Tech.
  {\bf 6} (2001) 74 [e-print arXiv:hep-th/0307230 (2003)].
\bibitem{SW2005} A.F.~Schunk and C.~Wainwright, {\it A geometric approach
  to scalar field theories on the supersphere}, J. Math. Phys.
  {\bf 46} (2005) 033511.
\bibitem{SNR19772}  M.~Scheunert, W.~Nahm and Y.~Rittenberg, {\it Irreducible
  representations of the \textit{osp}(2,1) and \textit{spl}(2,1) graded Lie
  algebras}, J. Math. Phys. {\bf 18} (1977) 155.
\bibitem{Hughes1981} J.W.~Hughes, {\it Representations of
  \textit{osp}(2,1) and the metaplectic representation}, J. Math. Phys.
  {\bf 22} (1981) 245.
\bibitem{Stewart1996} J.~Stewart, {\it Advanced general relativity},
  Cambridge University Press 1996.
\bibitem{Ilyenko2003} K.~Ilyenko, {\it Adjoint representation of the graded
  Lie algebra \textit{osp}(2/1;~$\mathbb{C}$) and its exponentiation}
  e-print arXiv:hep-th/0308009 (2003).
\bibitem{Cornwell1989} J.F.~Cornwell, {\it Group theory in physics}, Vol.~3,
  Academic Press 1989.
\bibitem{RS1978} V.~Rittenberg and M.~Scheunert, {\it Elementary construction of
  graded Lie groups}, J.~Math. Phys., \textbf{19} (1978) 709.
\bibitem{SNR19771}  M.~Scheunert, W.~Nahm and Y.~Rittenberg, {\it Graded Lie
  algebras: generalization of hermitian representations}, J. Math. Phys.
  {\bf 18} (1977) 146.
\bibitem{GK1985} L.E.~Gendenshtein and I.V.~Krive, {\it Supersymmetry in
  quantum mechanics}, Sov. Phys.~--~Usp. {\bf 28} (1985) 645.
\bibitem{MKS1966} W.~Magnus, A.~Karras and D.~Solitar, {\it
  Combinatorial group theory}, Interscience 1966.
\bibitem{Cartan1966} E.~Cartan, {\it The theory of spinors}, Hermann 1966.
\bibitem{Rashevskii1955} P.K.~Rashevskij, {\it The theory of spinors},
  Transl. Am. Math. Soc. {\bf II} (Ser.~6) (1957) 1.
\bibitem{DeWitt1992} B.~DeWitt, {\it Supermanifolds}, Cambridge Monographs on
  Mathematical Physics, CUP 1992.
\bibitem{KW2006} A.F.~Kleppe and C.~Wainwright, {\it Graded Majorana spinors},
  J.~Phys.~A {\bf 39} (2006) 3787.
\end{thebibliography}
\end{document}